\documentclass[aps,floatfix,superscriptaddress,twocolumn,showpacs,pra]{revtex4}
\usepackage{graphicx,bm,color}
\usepackage{array}
\usepackage{amsmath,amsfonts,amssymb}
\usepackage{pdflscape}
\usepackage{hyperref}
\hypersetup{colorlinks=true}
\hypersetup{urlcolor=blue}
\hypersetup{citecolor=black}
\hypersetup{menucolor=black}
\hypersetup{linkcolor=black}
\usepackage{cleveref}
\usepackage{longtable}
\usepackage{float}
\usepackage{pdfpages}
\usepackage[T1]{fontenc}
\usepackage[utf8]{inputenc}
\usepackage{bm}
\usepackage{ulem}

\newcommand{\be}{\begin{equation}}
\newcommand{\ee}{\end{equation}}
\newcommand{\bmul}{\begin{multline}}
\newcommand{\emul}{\end{multline}}
\newcommand{\bea}{\begin{eqnarray}}
\newcommand{\eea}{\end{eqnarray}}

\newcommand{\bra}[1]{\langle #1|}
\newcommand{\ket}[1]{|#1\rangle}

\newcommand{\meanv}[1]{\langle #1 \rangle}
\newcommand{\meanvlr}[1]{\left\langle #1 \right\rangle}

\newcommand{\bb}[1]{\left( #1 \right)}

\newcommand{\bbcro}[1]{\left[ #1 \right]}

\newcommand{\bbaco}[1]{\left\{ #1 \right\}}
\newcommand{\bbv}[1]{\left\vert #1 \right\vert}

\newcommand{\rr}{\textbf{r}}
\newcommand{\g}[1]{``#1''}

\newcommand{\ii}{\textrm{i}}
\newcommand{\eee}{\textrm{e}}
\newcommand{\dd}{\mathrm{d}}
\newcommand{\grad}{\vec{\nabla}}

\newcommand{\revision}[1]{#1}
\newcommand{\resoumis}[1]{#1}
\newcommand{\revisionbis}[1]{#1}
\newcommand{\elimination}[1]{#1}

\begin{document}
\title{EPR-entangled Bose-Einstein condensates in state-dependent potentials: a dynamical study}
\author{Hadrien Kurkjian}
\affiliation{\revisionbis{TQC, Universiteit Antwerpen, Universiteitsplein 1, B-2610 Antwerpen, Belgi\"e}}
\affiliation{\revisionbis{Laboratoire Kastler Brossel, ENS-PSL, CNRS, UPMC-Sorbonne Universit\'es and Coll\`ege de France, 24 rue Lhomond, 75231 Paris Cedex 05, France}}
\author{Krzysztof Paw\l owski}
\affiliation{\revisionbis{Center for Theoretical Physics, Polish Academy of Sciences, Al. Lotnik\'ow 32/46, 02-668 Warsaw, Poland}}
\author{Alice Sinatra}
\affiliation{\revisionbis{Laboratoire Kastler Brossel, ENS-PSL, CNRS, UPMC-Sorbonne Universit\'es and Coll\`ege de France, 24 rue Lhomond, 75231 Paris Cedex 05, France}}

\begin{abstract}
We study generation of non-local correlations by atomic interactions in a pair of bi-modal Bose-Einstein Condensates in state-dependent potentials including spatial dynamics.
The wave-functions of the four components are described by combining a Fock state expansion with a time-dependent Hartree-Fock Ansatz, so that both the spatial dynamics and the local and non-local quantum correlations are accounted for.
\resoumis{We find that despite the spatial dynamics, our protocole generates enough non-local entanglement to perform an {EPR steering experiment} with two spatially separated condensates of a few thousands of atoms.}
\end{abstract}


\maketitle

\section*{Introduction}
The non-local and non-deterministic nature of quantum mechanics is a subject of experimental investigation since a few decades. The most significant result in this direction was the violation of the so-called Bell inequalities with entangled pairs of photon \cite{Aspect1982}. With all loopholes now filled in \cite{Hanson2015}, this experiment ruled out the possibility that the phenomena described by quantum mechanics could by explained by an underlying local deterministic ``hidden variable'' theory. More recently, the observation of non-local entanglement between macroscopic massive objects has emerged as a new objective. This result would be an important step in further ruling out local realism, and in pushing back the boundary of observation of quantum effects in the macroscopic world. Experimentally, this is a challenging objective for two reasons: $(i)$ large systems decohere usually faster than individual systems and therefore retain their non-local entanglement only \revision{for} a short amount of time, $(ii)$ many-body systems have a complex internal dynamics which is likely to affect their non-local entanglement.

Cold atoms experiments provide a promising platform to tackle these limitations and observe non-local entanglement between reasonably large objects \resoumis{\cite{Treutlein2016}}. They provide experimentalists with clean isolated systems where the impact of decoherence can be fought more efficiently than in other many-body systems ; in particular, the dominant decoherence effect, atomic losses, can be made very weak for a careful choice of the atomic states and
can be accurately evaluated theoretically \cite{CastinSinatra2008}.
At low temperatures, one can prepare a system of bosonic atoms into a few macroscopically occupied modes, a situation which enables a natural generalization of the bipartite system imagined in a Bell-experiment to a macroscopic system.
Finally, the van der Walls interactions between the atoms are a powerful source of entanglement creation, able to produce highly entangled many-body states at short times, when the decoherence effects are limited. 

The correlations between the particles in a condensate of weakly-interacting bosons have recently been shown to be strong enough \cite{Sangouard2016} to allow in principle for a violation of multipartite Bell inequality \cite{Acin2014}. However a proper violation would require to perform measurements on each pair separately, which \revisionbis{is experimentally challenging}. Instead, \resoumis{it was recently proposed} \cite{Drummond2011,Kurizki2011} to violate a weaker form of non-locality usually referred to as ``\resoumis{EPR steering}'', by reference to the seminal article \cite{Rosen1935} which introduced the EPR paradox. In an EPR situation, the measurements done by Alice on her half of an entangled non-local quantum state are shown to apparently ``steer'' Bob's second half of the state, eventually leading to a violation of the Heisenberg uncertainty relation which Bob's observables should satisfy if the quantum state were purely local, the so-called ``EPR inequality''  \resoumis{\cite{Reid1989,Doherty2007}}. Such an experiment shows that the non-deterministic nature of Quantum Mechanics (in particular the Heisenberg uncertainty relation) is incompatible with locality, thus ruling out all hidden-variable theories locally compatible with Quantum Mechanics.

Already a violation of an EPR inequality \cite{Doherty2007} was obtained in a condensate of weakly-interacting bosons using state-changing collisions to create entangled subsystems of a few atoms \cite{Klempt2015}. An alternative route relying on light-matter interaction to create entanglement was explored in Ref.\cite{Polzik2001} and could be used in an EPR experiment.
Here we consider the entanglement scheme proposed by Refs.\cite{KPS2013} and \cite{Byrnes2013} where two Bose-Einstein condensates \resoumis{$a$ and $b$} of atoms in a superposition of two internal states \resoumis{$0$ and $1$} are entangled \textit{via} atomic interactions after a state-dependent transport (see Fig.\ref{fig:sequence}). \resoumis{As it was shown in Refs.\cite{KPS2013} and \cite{Byrnes2013}, this scheme allows to generate a variety of entangled states by unitary evolution with the Hamiltonian $\chi_{ab}\hat{S}_z^a\hat{S}_z^b$, where $\hat{S}_z^{a(b)}$ are the $z$ components of the collective spins of the two Bose-Einstein condensates $a$ and $b$. At remarquable times of the order of $\pi/\chi_{ab}$, one observes sharp dips in the entanglement entropy, associated with the formation of non-local macroscopic superpositions of coherent spin states which evoke the two-qubit Bell-state but in a $N$-body system. \resoumis{This Hamiltonian was recently realized with microwave cavities bridged by a superconducting artificial atom and used to generate entangled Schr\"odinger cat states of the electromagnetic field  \cite{Schoelkopf2016}.} The EPR correlations which are the focus of the present article appear on the other hand at much shorter times $1/N\ll \chi_{ab}t\ll1/\sqrt{N}$ \cite{KPS2013}. While no striking features appear in the entanglement entropy and phase space representation of the state at these times, the non-local correlations are large enough to violate an EPR steering inequality  \cite{Doherty2007} and the state, which can be seen as a non-local equivalent of a spin-squeezed state, is much more robust against decoherence.
{Using atomic interactions, spin-squeezed states have been already obtained in clouds whose size can reach a few thousand atoms \cite{KetterleSqueezing2007,Oberthaler2008,Treutlein2010}, and experiments along the same lines to demonstrate non-local entanglement could be envisaged.} }

However, fast transport of atomic clouds is likely to excite the spatial dynamics of the gas and one should check whether any non-local entanglement remains visible under these conditions. 
This is what we investigate in this paper by performing realistic numerical simulations of the intertwined spatial and entanglement dynamics. Our approach, adapted from Refs.\cite{CastinSinatra2000,TreutleinSinatra2009}, describes the state of the system as a coherent superposition of states with a fixed number of particles in a set of distinguishable modes (four in our case) representing the condensates. In each configuration, the wave functions evolve according to a time-dependent variational principle. This formalism is able to describe both the out-of-equilibrium dynamics of the gas, even with important excitations of the spatial eigenmodes, and the dynamics of entanglement between the distinguishable modes.
Using this powerful numerical approach, we investigate the effect of spatial dynamics on an EPR entanglement witness and provide experimental parameters for which a significant violation of an EPR inequality could be observed.


\section{The dynamical model}



\subsection{Time-dependent description of a system of multimode bosons}


\paragraph{Multimodal decomposition}

We consider a gas of bosonic atoms distributed in a number of modes of order unity. In practice, we will apply our model to the EPR situation of Ref.\cite{KPS2013} using four modes. We imagine a situation in which the atoms in different modes never have the chance to exchange and \revisionbis{can be considered as belonging to distinguishable components,} so that we can expand the bosonic field operator $\hat{\psi}$ in the form:
\be
\hat{\psi}(\rr)=\sum_{\alpha\in A} \hat{\psi}_{\alpha}(\rr)
\ee
where $\hat{\psi}_{\alpha}$ annihilates a boson in \revisionbis{component} $\alpha$ at site $\rr$, \revision{and $A$ is the set of distinguishable \revisionbis{components}.} We work with a space discretized into a cubic lattice of step $l$. This prefigures the numerical implementation of the model and removes any short-distance divergence. The commutation relations of the field operator are given by:
\be
[\hat{\psi}_{\alpha}(\rr),\hat{\psi}^\dagger_{\alpha'}(\rr')]=\frac{\delta_{\alpha,\alpha'}\delta_{\rr,\rr'}}{l^3}
\ee
We assume that the interactions between the atoms take place in the low-energy dilute regime usual in cold atoms, so that they can be represented by contact interactions. We note $g_{\alpha\alpha'}$ the coupling constant \revisionbis{between one
atom of the component $\alpha$ and one of $\alpha'$}. The Hamiltonian of this system reads:
\begin{multline}
\hat{H}=l^3\sum_{\rr,\revision{\alpha\in A}}\hat{\psi}^\dagger_\alpha(\rr) \bb{-\frac{\hbar^2}{2m}\Delta_\rr+V_\alpha(\rr,t)} \hat{\psi}_\alpha(\rr) \\
+ l^3 \sum_{\substack{\rr \\ \revision{\alpha,\alpha'\in A}}} \frac{g_{\alpha\alpha'}}{2} \hat{\psi}^\dagger_{\alpha}(\rr) \hat{\psi}^\dagger_{\alpha'}(\rr) \hat{\psi}_{\alpha}(\rr) \hat{\psi}_{\alpha'}(\rr)
\label{eq:hamiltonien} 
\end{multline}
\revisionbis{with the convention $g_{\alpha'\alpha}=g_{\alpha\alpha'}$.}
Note that we allow for a dependence of the external trapping potential $V_\alpha$ not only on time but also on the mode $\alpha$. This can be realized experimentally with state-dependent potentials \cite{Treutlein2010}.

\paragraph{Variational principle}

To describe the many-body dynamics arising from the Hamiltonian \eqref{eq:hamiltonien} we rely on a classical field theory based on the variational principle \cite{Ripka1985}. \revisionbis{We assume that for each distinguishable bosonic field $\hat{\psi}_\alpha$, there is one macroscopically populated single particle mode  $\phi_\alpha$ to be determined.} This wave function depends on position $\rr$, time $t$ and number of atoms $\vec{N}=(N_{\alpha'})_{\alpha'\in A}$ in each mode
\be
\phi_\alpha = \phi_\alpha (\vec{N},\rr,t), \qquad \vec{N}=(N_{\alpha'})_{\alpha'\in A}
\ee
We chose $\phi_\alpha$ to be normalized to unity:
\be
\forall\vec{N},\forall t, \quad l^3 \sum_{\rr} \bbv{ \phi_\alpha (\vec{N},\rr,t) }^2=1
\ee
Next, we introduce the operator which annihilates a boson in the wave function $\phi_\alpha$:
\be
\hat{a}_\alpha (\vec{N},t) \equiv l^3 \sum_{\rr} \phi_\alpha^* (\vec{N},\rr,t) \hat{\psi}_\alpha(\rr)
\ee
\revisionbis{This is a bosonic operator since $[\hat{a}_\alpha (\vec{N},t),\hat{a}_{\alpha'}^\dagger (\vec{N},t)]=\delta_{\alpha,\alpha'}$. We use it to form} the \g{Fock state} with $N_\alpha$ bosons in the wave function $\phi_\alpha$:
\be
\ket{\{N_\alpha:\phi_\alpha(\vec{N},\rr,t)\}} \equiv \frac{\prod_\alpha \bb{\hat{a}^\dagger_\alpha(\vec{N},t)}^{N_\alpha}}{\bb{\prod_\alpha N_\alpha!}^{1/2}} \ket{\rm vac}
\label{eq:etatFock} 
\ee
where $\ket{\rm vac}$ is the vacuum of bosons, \revision{and we use the short-hand notation $\sum_{\alpha}=\sum_{\alpha\in A}$, $\prod_{\alpha}=\prod_{\alpha\in A}$ and $\{X_\alpha\}=\{X_\alpha\}_{\alpha\in A}$}. Note that on state \eqref{eq:etatFock}, the field operator $\hat{\psi}_\alpha(\rr)$ acts as $\phi_\alpha(\vec{N},\rr,t) \hat{a}_\alpha(\vec{N},t)$. This can be seen by completing $\phi_\alpha$ into an orthonormal basis of the Hilbert space of mode $\alpha$ and by expanding the field operator over this basis.

We are now ready to apply the variational principle. The action between the initial time $t_{\rm i}$ and the final time $t_{\rm f}$ reads:
\begin{widetext}
\be
S \equiv \int_{t_{\rm i}}^{t_{\rm f}} \dd t \bbaco{\frac{\ii\hbar}{2} \bb{\bra{\{N_\alpha:\phi_\alpha(\vec{N},\rr,t)\}}\frac{\dd}{\dd t} \ket{\{N_\alpha:\phi_\alpha(\vec{N},\rr,t)\}} - {\rm c.c.} } -E(\vec{N},t)  }
\label{eq:variationnel} 
\ee
The first term between curly brackets in \eqref{eq:variationnel} can be expressed as
\be
\bra{\{N_\alpha:\phi_\alpha(\vec{N},\rr,t)\}}\frac{\dd}{\dd t} \ket{\{N_\alpha:\phi_\alpha(\vec{N},\rr,t)\}} - {\rm c.c.}  =  2l^3 \sum_{\rr,\alpha} N_\alpha \phi_\alpha^*(\vec{N},\rr,t) \frac{\dd}{\dd t} \phi_\alpha(\vec{N},\rr,t)
\label{eq:entreaccolades} 
\ee
and the energy $E(\vec{N},t)$ is given by
\begin{multline}
E(\vec{N},t) \equiv \bra{\{N_\alpha:\phi_\alpha(\vec{N},\rr,t)\}}\hat{H} \ket{\{N_\alpha:\phi_\alpha(\vec{N},\rr,t)\}} \\
= l^3\sum_{\rr,\alpha} N_\alpha \phi_\alpha^*(\vec{N},\rr,t) h_\alpha \phi_\alpha(\vec{N},\rr,t) + l^3 \sum_{\rr,\alpha} \frac{g_{\alpha \alpha }}{2} N_\alpha (N_\alpha-1) |\phi_\alpha(\vec{N},\rr,t)|^4  
+ l^3\!\!\!\!\!\! \sum_{\rr,\alpha,\alpha'\neq\alpha} \!\!\! \frac{g_{\alpha \alpha' }}{2} N_\alpha N_{\alpha'} |\phi_\alpha(\vec{N},\rr,t)|^2 |\phi_{\alpha'}(\vec{N},\rr,t)|^2
\label{eq:energie} 
\end{multline}
\end{widetext}
where we have defined
\be
h_\alpha(\rr,t) = {-\frac{\hbar^2}{2m}\Delta_\rr+V_\alpha(\rr,t)}
\ee

The extremalisation of the action, $\forall\rr,\forall t,\, \delta S/\delta \phi^*_\alpha(\vec{N},\rr,t)=0$, combined with Eqs.\eqref{eq:entreaccolades} and \eqref{eq:energie} leads to the equation of evolution of $\phi_\alpha$, also known as the \g{time-dependent Gross-Pitaevskii equation}:
\begin{widetext}
\be
\ii\hbar \frac{\dd\phi_\alpha(\vec{N},\rr,t)}{\dd t} = \bb{h_\alpha(\rr,t)  
+g_{\alpha \alpha} (N_\alpha-1) \bbv{  \phi_\alpha(\vec{N},\rr,t) }^2   
+ \sum_{\alpha'\neq\alpha} g_{\alpha \alpha'} N_{\alpha'} \bbv{  \phi_{\alpha'}(\vec{N},\rr,t) }^2  }\phi_\alpha(\vec{N},\rr,t)
\label{eq:gp_dept} 
\ee
\end{widetext}


\paragraph{Accumulated phase of a Fock state}


We now want to describe the evolution of the quantum state describing the gas. We imagine that initially this state coincides with a Fock state:
\be
\ket{\psi(0)} = \ket{\{N_\alpha:\phi_\alpha(\vec{N},\rr,0)\}}
\ee
\revision{Without approximations}, it is difficult to follow the evolution of this state because the wave functions $\phi_\alpha$ are not the exact eigenmodes of the Hamiltonian \eqref{eq:hamiltonien}, therefore the state after an evolution time $t$ is no longer a Fock state. We overcome this difficulty by assuming that the coupling to the modes orthogonal to $\phi_\alpha$ remains weak all along the evolution. This is justified as long as the dynamics does not create a significant depletion in the condensates. Under this assumption, the state at time $t$ can be approximated by an Hartree-Fock Ansatz:
\be
\eee^{-\ii\hat{H}t/\hbar}\ket{\psi(0)} \simeq \eee^{-\ii A(\vec{N},t)} \ket{\{N_\alpha:\phi_\alpha(\vec{N},\rr,t)\}}
\ee
where the accumulated phase $A(\vec{N},t)$ is self-consistently defined by
\begin{multline}
\eee^{-\ii A(\vec{N},t)} = \\ \bra{\{N_\alpha:\phi_\alpha(\vec{N},\rr,t)\}} \eee^{-\ii\hat{H}t/\hbar} \ket{\{N_\alpha:\phi_\alpha(\vec{N},\rr,0)\}}
\end{multline}
Deriving this relation with respect to time we obtain its equation of motion
\begin{widetext}
\be
 \hbar \frac{\dd A(\vec{N},t) }{\dd t} 
=-l^3\sum_{\rr,\alpha} \frac{g_{\alpha\alpha}}{2} N_\alpha (N_\alpha-1) \bbv{ \phi_\alpha(\vec{N},\rr,t) }^4  
 -l^3\sum_{\rr,\alpha,\alpha'\neq\alpha} \frac{g_{\alpha\alpha'}}{2} N_\alpha N_{\alpha'} \bbv{ \phi_\alpha(\vec{N},\rr,t) }^2 \bbv{ \phi_{\alpha'}(\vec{N},\rr,t) }^2
\label{eq:derivA} 
\ee
\end{widetext}
which completes our description of the dynamics of the gas.


\subsection{EPR experiment with two bimodal condensates}
\label{sec:coherentsuperposition}

\paragraph{Coherent superposition of Fock states}


\begin{figure}[t]
\centering
\includegraphics[width=0.3\textwidth]{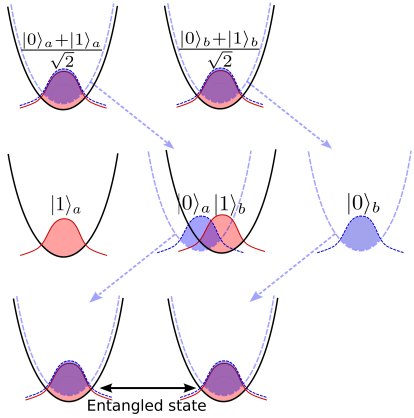}
\caption{\revision{(Color online) Sequence allowing to entangle condensate {\revisionbis{$a$  (left well) and $b$ (right well)}} via controlled {collisional} interaction in state dependent trapping potentials. The interaction phase, where  the correlations are created, 
is depicted in the central panel.}
\label{fig:sequence}}
\end{figure}

We now place ourselves in the experimental situation of Ref.\cite{KPS2013} (represented in Fig.\ref{fig:sequence}) where two initially uncorrelated Bose-Einstein condensates \revisionbis{of atoms in their internal ground state denoted $0$} are prepared in a double well potential \revisionbis{in an initial wave function $\phi_{a0}(\rr,t=0)$ for the $a$ (or left) well and $\phi_{b0}(\rr,t=0)$ for the $b$ (or right) well}. In this study, we assume that the number of particles is fixed to $N_a$ in the \revisionbis{$a$} well and to $N_b$ in the \revisionbis{$b$} well. In each well, an electromagnetic pulse brings the atoms in a superposition of two internal states noted \revision{${0}$ and ${1}$}. We assume that the four distinguishable \revision{components} thus formed
\be
\alpha=\sigma,\epsilon \qquad \mbox{with} \qquad \sigma=a,b \quad \mbox{and} \quad \epsilon=0,1
\label{eq:4modes} 
\ee
will remain so during the entire evolution. This is the case in the scheme proposed in Fig.\ref{fig:sequence}.
Just after the pulses which prepares a coherent superposition of internal states $0$ and $1$ in each well, the state is $a$-$b$-factorized and can be written in the compact form:
\bea
\!\!\!\!\!\!\!\!\!\!\ket{\Psi(0)}
\!\!\!&=&\!\!\revisionbis{\prod_{\sigma=a,b} \frac{(C_{\sigma,0} \hat{a}_{\sigma,0}(0)^\dagger+C_{\sigma,1} \hat{a}_{\sigma,1}(0)^\dagger)^{N_\sigma} }
{(N_\sigma!)^{1/2}} \ket{\rm vac}} \\
\!\!\!\!&=&\!\!\!\sum_{\vec{N}\in\mathcal{N}}  
\bb{\frac{N_a ! N_b !}{\prod_{\alpha} N_{\alpha}!}}^{1/2}
\prod_{\alpha} C_{\alpha}^{N_{\alpha}} \left\vert\bbaco{N_{\alpha}:\phi_{\alpha}(\rr,0)}\right\rangle
\label{eq:psi_decompo} 
\eea
where the coefficients $C_\alpha$ depend on the strength \revision{and on the phase of the pulses} (for two $\pi/2$-pulses, $\forall\alpha,\, C_\alpha=1/\sqrt{2}$) and the atom number conservation in each well constrains the vector $\vec{N}=(N_\alpha)_{\alpha\in A}$ to belong to the subset of $\mathbb{N}^4$:
\be
\mathcal{N}=\{\vec{N} \in \mathbb{N}^4, N_{a,0}+N_{a,1}=N_a,\: \text{and}\: N_{b,0}+N_{b,1}=N_b \}
\ee
\revisionbis{The operators $\hat{a}_{\sigma,0}^\dagger(0)$ and $\hat{a}_{\sigma,1}^\dagger(0)$ create atoms in the same wave functions $\phi_{\sigma0}(\rr,t=0)=\phi_{\sigma1}(\rr,t=0)$, which do not depend on the population $\vec{N}\in\mathcal{N}$, hence the omission of this variable in the list of arguments.}
The decomposition \eqref{eq:psi_decompo} suggests a two-fold naive procedure to compute the state $\ket{\Psi(t)}$ of the gas at time $t$:
\begin{enumerate}
\item we compute the evolved wave function $\phi_\alpha(\vec{N},\rr,t)$ for each mode $\alpha$ and for all vectors $\vec{N}$ such that the prefactor $\bb{{N_a ! N_b !}/{\prod_{\alpha} N_{\alpha}!}}^{1/2} \prod_{\alpha} C_{\alpha}^{N_{\alpha}}$ is significantly non-zero. In the case of $\pi/2$-pulses this prefactor is maximal for $\vec{\bar{N}}=(N_a/2,N_a/2,N_b/2,N_b/2)$ and decreases around this maximum with a typical width $N_{a,0}-N_a/2=O(N_a^{1/2})$ and $N_{b,0}-N_b/2=O(N_b^{1/2})$. With this procedure, a typical number of order $O(N_a^{1/2} N_b^{1/2})$ of wave functions must be computed and evolved.
\item we evolve the phase factor $A(\vec{N},t)$ using Eq.\eqref{eq:derivA}, again for all the values of $\vec{N}$ which contribute significantly to the superposition \eqref{eq:psi_decompo}.
\end{enumerate}
Doing so, we obtain the state
\begin{multline}
\ket{\Psi(t)}=\sum_{\vec{N}\in\mathcal{N}}  
\bb{\frac{N_a ! N_b !}{\prod_{\alpha} N_{\alpha}!}}^{1/2}
\eee^{-\ii A(\vec{N},t)} \\
\times \prod_{\alpha} C_{\alpha}^{N_{\alpha}} \left\vert\bbaco{N_{\alpha}:\phi_{\alpha}(\vec{N},\rr,t)}\right\rangle
\label{eq:psi_t} 
\end{multline}
Although analytically attractive, this naive procedure seems difficult to implement numerically for large atom numbers for two reasons: $(i)$ it requires us to compute and store a \elimination{senselessly} large number of wave functions as soon as $N_a,N_b$ reach mesoscopic values and $(ii)$ it supposes that we calculate the phase factor $A$, which is of order $O(N)$ (see Eq.\eqref{eq:derivA}), with an accuracy of order $O(1/N)$ \revisionbis{as it will be clear in section \ref{sec:modulephase}.} \revision{In that section, we will explain how to overcome these difficulties with the help of the modulus-phase approximation. Before we do so, let us explain in the following paragraph how a generalized EPR experiment can be performed in our system of multimode bosons.}

\paragraph{Collective spins and entanglement witness}

To perform a generalized EPR experiment on our many-body system, we need to define generalized conjugated variables in each subsystem $a$ and $b$. To this end, we view each two-level atom as an effective spin and define the collective spins 
\bea
\hat{S}_x^\sigma &\equiv& \frac{1}{2}\ l^3 \sum_\rr \bb{\hat{\psi}_{\sigma 0}^\dagger(\rr) \hat{\psi}_{\sigma 1}(\rr) + \hat{\psi}_{\sigma 1}^\dagger(\rr) \hat{\psi}_{\sigma 0}(\rr) } \\
\hat{S}_y^\sigma &\equiv& \frac{\ii}{2}\ l^3 \sum_\rr \bb{\hat{\psi}_{\sigma 0}^\dagger(\rr) \hat{\psi}_{\sigma 1}(\rr) - \hat{\psi}_{\sigma 1}^\dagger(\rr) \hat{\psi}_{\sigma 0}(\rr) } \\
\hat{S}_z^\sigma &\equiv& \frac{1}{2}\ l^3 \sum_\rr \bb{\hat{\psi}_{\sigma 1}^\dagger(\rr) \hat{\psi}_{\sigma 1}(\rr) - \hat{\psi}_{\sigma 0}^\dagger(\rr) \hat{\psi}_{\sigma 0}(\rr) }
\eea
which form a spin algebra in each well $\sigma=a,b$. The average spins precess around the $z$ direction at an angular frequency given by the chemical potential difference between states $0$ and $1$. We get rid of this effect by unrotating the spin in the $x,y$ plane
\bea
\hat{S}_{x\phi}^\sigma &=& \cos \phi_\sigma \hat{S}_{x}^\sigma + \sin \phi_\sigma \hat{S}_{y}^\sigma \\
\hat{S}_{y\phi}^\sigma &=&-\sin \phi_\sigma \hat{S}_{x}^\sigma + \cos \phi_\sigma \hat{S}_{y}^\sigma 
\eea
where the rotation angle is defined from the average spin
\be
\mbox{tan}\phi_\sigma = \frac{\meanv{\hat{S}_{y}^\sigma}}{\meanv{\hat{S}_{x}^\sigma}}
\ee
The interesting entanglement dynamics takes place in the $y\phi,z$ plane. Because of the non-linearity of the interaction term of the Hamiltonian \eqref{eq:hamiltonien}, non-classical correlations build up between the spin components $\hat{S}_{y\phi}^a$, $\hat{S}_{z}^a$, $\hat{S}_{y\phi}^b$, and $\hat{S}_{z}^b$. The correlations within one well are responsible for spin-squeezing ; they are not our focus in this article. We are looking for the highest non local correlations between components of the $a$ and $b$ wells. These are obtained for some well chosen quadratures in each well
\bea
\hat{S}_\alpha^a &=&\cos \alpha\ \hat{S}_{y\phi}^{a}   + \sin\alpha\ \hat{S}_z^{a} \label{eq:alpha} \\
\hat{S}_\beta^b &=&\cos \beta\ \hat{S}_{y\phi}^{b}       + \sin\beta\ \hat{S}_z^{b} 
\label{eq:beta}
\eea
Our EPR entanglement witness \cite{Drummond2012,KPS2013} is constructed from these optimally correlated quadratures and their conjugated ones
\begin{widetext}
\begin{equation}
E_{\rm EPR}^2 \equiv \frac{ 4 \bb{\Delta^2 \hat{S}_\alpha^a \Delta^2 \hat{S}_\beta^b-\text{Covar}^2(\hat{S}_\alpha^a,\hat{S}_\beta^b)}\bb{\Delta^2 \hat{S}_{\alpha+\pi/2}^a \Delta^2 \hat{S}_{\beta+\pi/2}^b-\text{Covar}^2(\hat{S}_{\alpha+\pi/2}^a,\hat{S}_{\beta+\pi/2}^b)}}{\bb{\Delta^2\hat{S}_\alpha^a\Delta^2\hat{S}_{\alpha+\pi/2}^a}  \left|\meanv{\hat{S}_{x\phi}^b}\right|^2 }
\label{eq:E_EPR}
\end{equation}
\end{widetext}
When this quantity is below unity, it shows a violation of the Heisenberg relation which Bob should expect if his quantum state were local, namely:
\be
\Delta^2(\hat{S}_{\beta}^b-\hat{S}_{\beta}^{b,{\rm inf}}) \Delta^2(\hat{S}_{\beta+\pi/2}^b-\hat{S}_{\beta+\pi/2}^{b,{\rm inf}})  < \frac{1}{4} \left|\meanv{\hat{S}_{x\phi}^b}\right|^2
\label{eq:EPRcriterion}
\ee
where $\Delta^2\hat{S}$ is the variance of $\hat{S}$ and $\hat{S}_{\beta}^{b,{\rm inf}}$ is the best guess that Alice can do on the outcome of the measurement of $\hat{S}_{\beta}^{b}$, knowing the outcome of her measurement of $\hat{S}_{\alpha}^{a}$:
\be
\hat{S}_{\beta}^{b,{\rm inf}}=\meanv{\hat{S}_{\beta}^{b}}+\frac{\mbox{Covar}(\hat{S}_{\alpha}^{a},\hat{S}_{\beta}^{b})}{\Delta^2 \hat{S}_{\alpha}^{a}} \bb{\hat{S}_{\alpha}^{a}-\meanv{\hat{S}_{\alpha}^{a}}}
\ee
\resoumis{This is the so-called ``steering'' effect: the violation of the Heisenberg inequality shows that Bob's measurement acts instantaneously and at a distance on the quantum state of Alice. If one obtains a value of $E_{\rm EPR}$ below unity, one not only proves the entanglement between the $a$ and $b$ subsystems but also rules out the possibility of reconciling Quantum Mechanics (namely the Heisenberg uncertainty principle) with local realism \cite{Doherty2007,KPS2013}. However, $E_{\rm EPR}<1$ does not prove that no hidden variable theory can account for the results of the experiment, and is therefore strictly weaker than the violation of a Bell inequality.}

\subsection{Modulus-phase approximation}
\label{sec:modulephase}
We now explain how to overcome the limitation mentioned in \S\ref{sec:coherentsuperposition}.$a$ by performing a linearization of the phase of the wave functions around the central Fock state.

\paragraph{Modulus-phase decomposition of the wave functions}

The modulus-phase approximation consists in a linearization of $\phi_\alpha(\vec{N})$ around its central \elimination{and average} value for $\vec{N}=\vec{\bar{N}}$, where $\vec{\bar{N}}$ is the configuration that has the greatest weight in the superposition \eqref{eq:psi_decompo}. This approximation further assumes that the variation of the modulus of $\phi_\alpha$ can be entirely neglected \cite{CastinSinatra2000,TreutleinSinatra2009}.
It is valid when $N_a,N_b$ are large enough \revisionbis{such that} the typical width of the distribution of Fock states in state \eqref{eq:psi_decompo} (which is of order $O(\sqrt{N})$) \revisionbis{is} small compared to the central values $\bar{N}_{a,0},\bar{N}_{b,0}$. We write the wave function $\phi_\alpha$ for $\vec{N}$ such that $N_\alpha-\bar{N}_\alpha=O(N^{1/2})$ in the form:
\bea
\phi_\alpha (\vec{N},\rr,t) &\simeq& \bbv{\bar{\phi}_\alpha(\rr,t)} \eee^{\ii\theta_\alpha(\vec{N},\rr,t)} \\
&=& \bar{\phi}_\alpha(\rr,t) \eee^{\ii (\vec{N}-\vec{\bar{N}})\cdot\vec{\nabla} \theta_\alpha (\vec{\bar{N}},\rr,t) \revisionbis{+ O\bb{\frac{1}{N}}}} \notag
\eea
where
\be
 \bar{\phi}_\alpha(\rr,t)\equiv\phi_\alpha (\vec{\bar{N}},\rr,t)= \bbv{\bar{\phi}_\alpha(\rr,t)} \eee^{\ii\theta_\alpha (\vec{\bar{N}},\rr,t)}
\ee
is the central wave function of the mode $\alpha$, and the vector $\vec{\nabla}$ gathers the derivatives with respect to the number of atoms in each mode:
\be
\vec{\nabla} = \bb{ \frac{\partial}{\partial N_\alpha} \Big\vert_{\rr,t,N_{\alpha'\neq\alpha}} }_{\alpha\in A}
\ee
Thanks to this approximation, we need only the central wave functions $\bar{\phi}_\alpha$ and the derivatives $\vec{\nabla}\theta_\alpha$ of the phases in $\vec{N}=\vec{\bar{N}}$ to describe the state $\ket{\Psi(t)}$ of the gas. In practice, we compute $\vec{\nabla}\theta_\alpha$ using, besides the central wave function $\bar{\phi}_\alpha$, the slightly displaced wave functions $\phi_\alpha(\vec{\bar{N}}+\vec{\beta})$ where $\vec{\beta}=(\pm\beta_\alpha)_{\alpha\in A}$ and $\beta_\alpha\ll\bar{N}_\alpha$. In the $4$-mode situation considered in \eqref{eq:4modes}, this means evolving a total of only nine different Fock states, hence storing only $36$ wave functions.


\paragraph{Simplification of the accumulated phase}
\label{sec:compensation}

The objective of this section is to \revisionbis{show how the computation of the accumulated phase $A$ can be avoided}. In practice, rather than with the factor $A$ directly, we deal with differences of the kind $A(\vec{N}+\vec{\beta}) - A(\vec{N})$, where $\vec{\beta}$ represents a displacement of order unity of the population of each mode. We intend to show that this factor can be reabsorbed by a product of wave function overlaps. The quantity which appears naturally when one computes the averages of the spin components (see below Sec. \ref{sec:moyennes} and in particular Eqs.\eqref{eq:resultat_intermediaire} and \eqref{eq:recouvrement}) is the following combination:
\begin{widetext}
\be
\Theta(\vec{N},\vec{\beta},t) = \hbar \bb{A(\vec{N}+\vec{\beta},t) - A(\vec{N},t)} 
 - \ii\hbar \sum_\alpha \bb{N_\alpha+\frac{\beta_\alpha-1}{2}} \ln \bbcro{l^3 \sum_\rr \phi_\alpha^*(\vec{N}+\vec{\beta},\rr,t) \phi_\alpha(\vec{N},\rr,t)}
\label{eq:defTheta}
\ee
\end{widetext}
In contrast with $A$, the phase $\Theta$ has a small time derivative and is therefore easy to compute numerically. We give the real part of this time-derivative in the modulus-phase approximation:
\begin{multline}
\!\!\!\mbox{Re}\,\frac{\dd}{\dd t} \Theta(\vec{N},\vec{\beta},t) = -l^3\!\!\!\!\sum_{\rr,\alpha,\alpha'\neq\alpha} \!\!\!\!\frac{g_{\alpha\alpha'}}{2} \beta_{\alpha'} | \bar{\phi}_\alpha(\rr,t)|^2 |\bar{\phi}_{\alpha'}(\rr,t)|^2 \\ + O\bb{\frac{1}{N^2}}=O\bb{\frac{1}{N}}
\label{eq:dThetadt}
\end{multline}
\revisionbis{To obtain this result, we use Eq.\eqref{eq:derivA} to express the time-derivative of the first term between parenthesis in Eq.\eqref{eq:defTheta}, and the Gross-Pitaevskii equation \eqref{eq:gp_dept} to express the time-derivative of the second term.} Note that, consistently with the fact that we neglected the variations of the modulus of $\phi_\alpha$, we neglect the imaginary part of $\dd\Theta/\dd t$ which (like its real part) is of order $O(1/N)$.

\subsection{Quantum averages of field operators}
\label{sec:moyennes}
We conclude this section by deriving a general formula for the \revisionbis{two-position densities of the quantum averages}. Let us write an arbitrary products of field operators taken at different sites $\rr$ et $\rr'$ as:
\be
\hat{W}_{\substack{\vec{\gamma},\vec{\gamma}' \\ \vec{\delta},\vec{\delta'}}}(\rr,\rr')=\prod_{\alpha} (\hat{\psi}_{\alpha}^\dagger(\rr))^{\gamma_{\alpha}} (\hat{\psi}_{\alpha}^\dagger(\rr'))^{\gamma'_{\alpha}} (\hat{\psi}_{\alpha}(\rr))^{\delta_{\alpha}} (\hat{\psi}_{\alpha}(\rr'))^{\delta'_{\alpha}}
\label{eq:operateur_moyenne} 
\ee
where the vectors $\vec{\gamma},\vec{\gamma'},\vec{\delta},\vec{\delta'}\in\mathbb{N}^4$ are of order unity. For the sake of readability, we shall use the short-hand notations
\be
\vec{\delta}^+=\vec{\delta}+\vec{\delta}' \qquad \mbox{and} \qquad  \vec{\gamma}^+=\vec{\gamma}+\vec{\gamma}'
\ee
\revision{With the knowledge of $\meanv{\hat{W}(\rr,\rr')}$ for all $\vec{\gamma},\vec{\gamma'},\vec{\delta},\vec{\delta'}$ in hand, all the spin averages needed to compute $E_{\rm EPR}$ in Eq.\eqref{eq:E_EPR} can be reconstructed by integrating over $\rr$ and $\rr'$ and doing simple linear combinations.}
We compute the average value of \revisionbis{$\hat{W}(\rr,\rr')$} using the expansion \eqref{eq:psi_t} of $\ket{\Psi(t)}$ over the Fock states:
\begin{widetext}
\begin{multline}
\meanvlr{\hat{W}_{\vec{\gamma},\vec{\gamma'},\vec{\delta},\vec{\delta'}}(\rr,\rr')} = \sum_{\vec{N},\vec{N'}\in\mathcal{N}} \bb{\frac{N_a!^2N_b!^2}{\prod_\alpha (N_\alpha-\delta_\alpha^+)! (N_\alpha'-\gamma_\alpha^+)!}}^{1/2} \eee^{\ii\bb{A(\vec{N}',t)-A(\vec{N},t)}} \\
\times \prod_\alpha \bbcro{(C_\alpha^*)^{N_\alpha'} C_\alpha^{N_\alpha}  \bb{\phi_{\alpha}^*(\vec{N}',\rr,t)}^{\gamma_\alpha} \bb{\phi_{\alpha}^*(\vec{N}',\rr',t)}^{\gamma'_\alpha}   \bb{\phi_{\alpha}(\vec{N},\rr,t)}^{\delta_\alpha} \bb{\phi_{\alpha}(\vec{N},\rr',t)}^{\delta'_\alpha} } \\
\times \left\langle \bb{N_{\alpha}'-\gamma_\alpha^+:\phi_{\alpha}(\vec{N}',\rr'',t)}_\alpha \right\vert \left.  \bb{N_{\alpha}-\delta_\alpha^+:\phi_{\alpha}(\vec{N},\rr'',t)}_\alpha \right\rangle
\label{eq:resultat_intermediaire} 
\end{multline}
The scalar product appearing on the last line of \eqref{eq:resultat_intermediaire} is non-zero if and only if $\vec{N}'=\vec{N}+\vec{\gamma}^+-\vec{\delta}^+$ ; it can be computed without difficulty:
\begin{multline}
\!\!\!\!\!\!\!\!\!\left\langle \bb{N_{\alpha}-\delta_\alpha^+:\phi_{\alpha}(\vec{N}+\vec{\gamma}^+ - \vec{\delta}^+,\rr,t)}_\alpha \right\vert \! \left.  \bb{N_{\alpha}-\delta_\alpha^+:\phi_{\alpha}(\vec{N},\rr,t)}_\alpha \right\rangle \! = \! \prod_\alpha \bbcro{l^3 \sum_\rr \phi_\alpha^*(\vec{N}+\vec{\gamma}^+-\vec{\delta}^+,\rr,t) \phi_\alpha(\vec{N},\rr,t)}^{N_\alpha-\delta_\alpha^+}
\label{eq:recouvrement}
\end{multline}
\end{widetext}
We now use the modulus-phase approximation in order to $(i)$ compute the product of wave functions appearing on the second line of \eqref{eq:resultat_intermediaire} and to $(ii)$ eliminate the phase factor $A(\vec{N}+\vec{\gamma}^+-\vec{\delta}^+) - A(\vec{N})$ \revisionbis{in favor of the quantity $\Theta(\vec{N},\vec{\gamma}^+-\vec{\delta}^+,t)$ defined in Eq.\eqref{eq:defTheta}.} We obtain the slightly long but very general formula
\begin{widetext}
\begin{multline}
\meanvlr{\hat{W}_{\vec{\gamma},\vec{\gamma'},\vec{\delta},\vec{\delta'}}(\rr,\rr')} \underset{\mbox{modulus-phase}}{=} \sum_{\vec{N}\in\mathcal{N}} {\frac{N_a!N_b!}{\prod_\alpha (N_\alpha-\delta_\alpha^+)!}}  \\ \times \prod_\alpha  \bbcro{|C_\alpha|^{2N_\alpha} (C_\alpha^*)^{\gamma_\alpha^+ - \delta_\alpha^+}  \bb{\bar{\phi}_{\alpha}^*(\rr,t)}^{\gamma_\alpha} \bb{\bar{\phi}_{\alpha}^*(\rr',t)}^{\gamma'_\alpha}   \bb{\bar{\phi}_{\alpha}(\rr,t)}^{\delta_\alpha} \bb{\bar{\phi}_{\alpha}(\rr',t)}^{\delta'_\alpha} } \\
\times \eee^{\ii \bb{\vec{N}-\vec{\bar{N}}} \cdot \grad \bb{(\vec{\delta}-\vec{\gamma})\cdot\vec{\theta} (\rr,t)  + (\vec{\delta}'-\vec{\gamma}')\cdot\vec{\theta} (\rr',t) }} 
 \eee^{\ii \bb{\vec{\delta}^+ - \vec{\gamma}^+} \cdot \grad \bb{\vec{\gamma}\cdot\vec{\theta} (\rr,t)  + \vec{\gamma}' \cdot\vec{\theta} (\rr',t) }} \\
\times \exp\bbcro{\frac{\ii \Theta(\vec{N},\vec{\gamma}^+ - \vec{\delta}^+,t)}{\hbar} - \sum_\alpha \frac{\delta_\alpha^+ + \gamma_\alpha^+ -1}{2} \ln \bb{l^3\sum_\rr \phi_\alpha^*(\vec{N}+\vec{\gamma}^+ - \vec{\delta}^+,\rr,t) \phi_\alpha^*(\vec{N},\rr,t) } }
\label{eq:resultat_final} 
\end{multline}
\end{widetext}
where $\vec{\theta} = (\theta_\alpha)_{\alpha\in A}$.
In the numerical simulation, all the relevant average values can be calculated from this formula after a double summation over the spatial indices $\rr$ and $\rr'$. This result generalizes and gives a compact form
to the equations of appendix B of Ref.\cite{TreutleinSinatra2009}.

\section{Numerical simulation}
\label{sec:numerical}

\begin{figure*}[htb]
\begin{minipage}[l]{0.47\linewidth}
$(a)$ $N_a=N_b=100$ \\
\includegraphics[width=\textwidth]{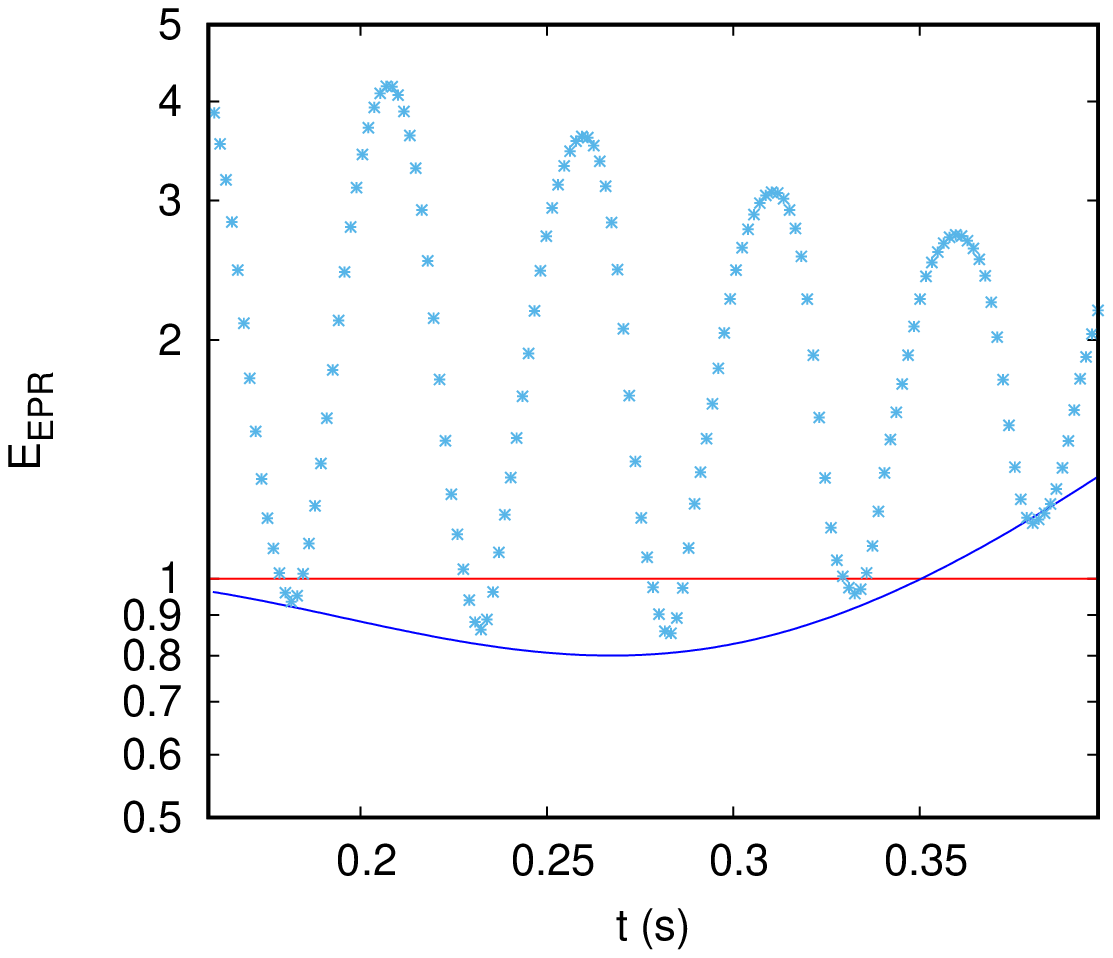} \\
\includegraphics[width=\textwidth]{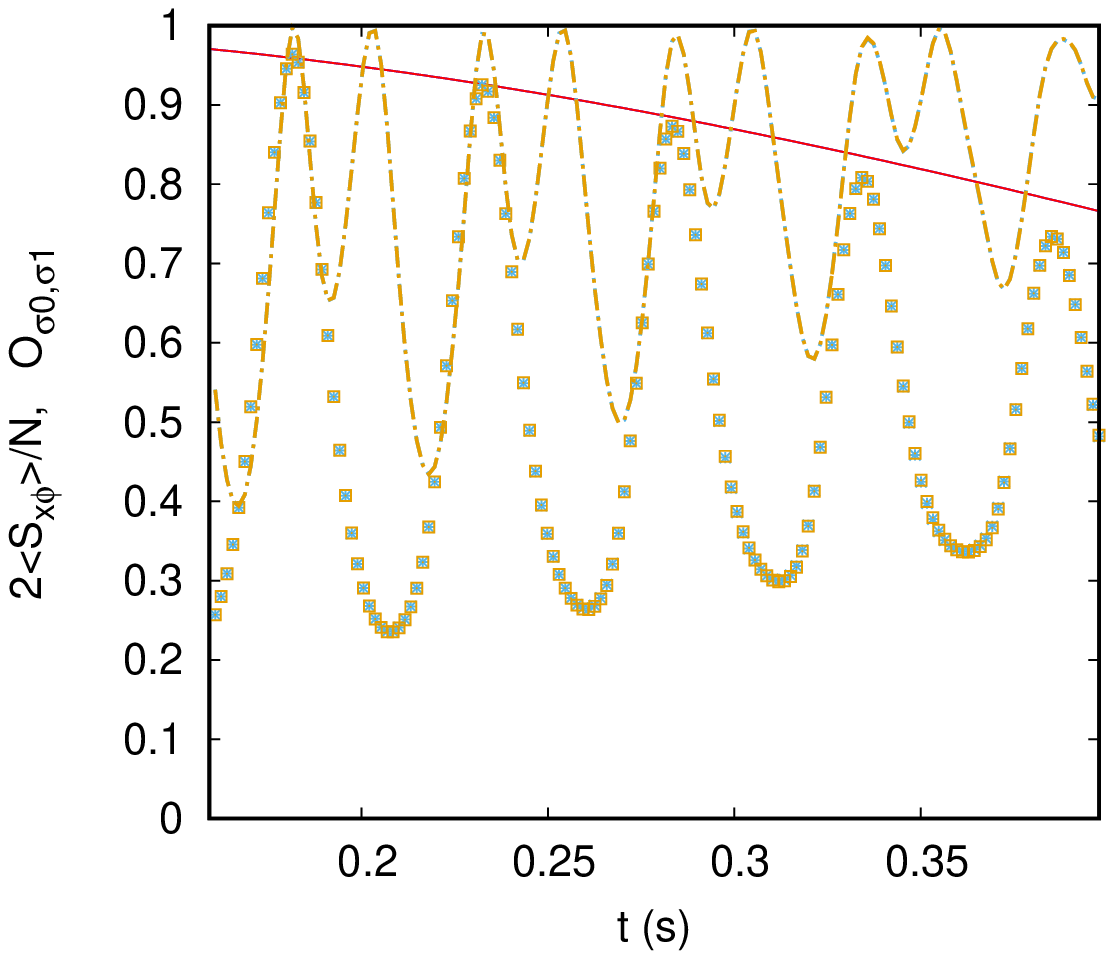}
\end{minipage}
\begin{minipage}[c]{0.47\linewidth}
$(b)$ $N_a=N_b=500$ \\
\includegraphics[width=\textwidth]{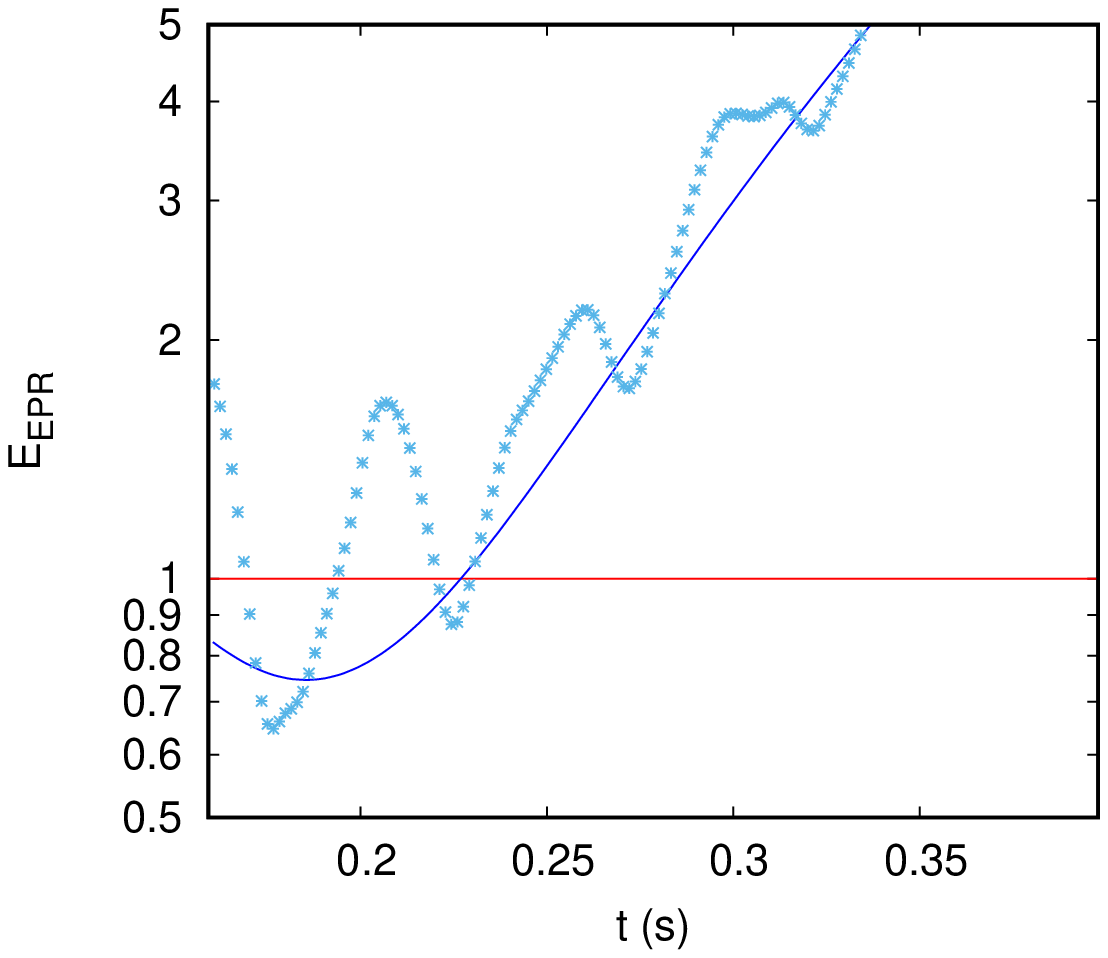}\\
\includegraphics[width=\textwidth]{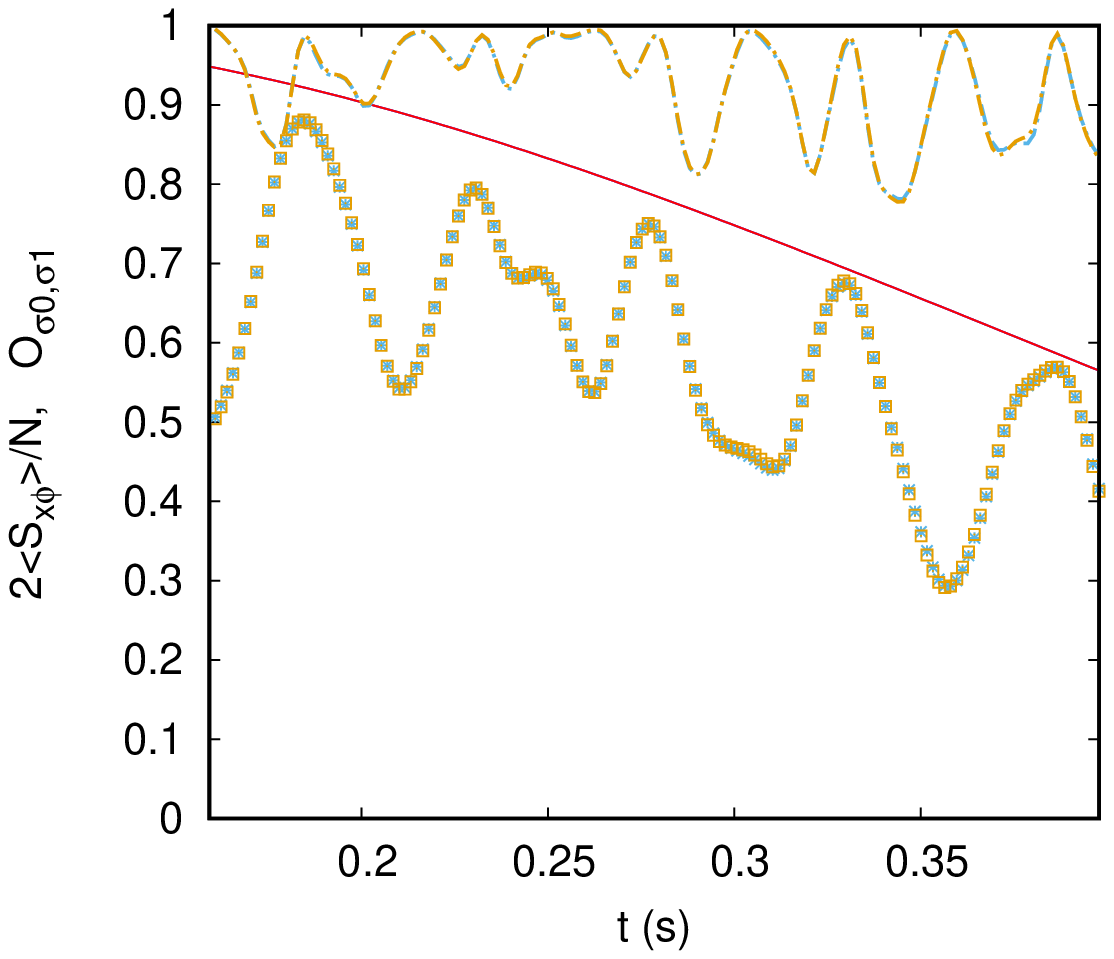}
\end{minipage}
\caption{(Color online) The EPR entanglement witness $E_{\rm EPR}$ (top row) and the renormalized lengths of the two spins $2\meanv{\hat{S}_{x\phi}^a}/N_a$ and  $2\meanv{\hat{S}_{x\phi}^b}/N_b$ (bottom row) are plotted as a function of the total evolution time $t=2t_{\rm R}+t_{\rm int}$ for (a) $N_a=N_b=100$ and $(b)$ $N_a=N_b=500$. The trap frequency is $\omega=2\pi\times 20\, \mbox{Hz}$ and the scattering lengths \resoumis{$a_{00}=100.4\, R_{\rm B}$, $a_{11}= 95.0\, R_{\rm B}$ and $a_{01}=98.0\, R_{\rm B}$} in units of the Bohr radius. The initial separation between the traps is $\delta z_{\rm max}=10 a_0$ in unit of the oscillatory length $a_0=\sqrt{\hbar/m \omega}$ and the ramping time \resoumis{is} $\omega t_{\rm R}=10$. Symbols: result of numerical simulations of spatial dynamics in the modulus-phase approximation (see Sec \ref{sec:modulephase}). \resoumis{In the top row, the blue crosses represent $E_{\rm EPR}$. In the bottom row, the blue crosses and yellow squares are for $\meanv{\hat{S}_{x\phi}^a}$ and $\meanv{\hat{S}_{x\phi}^b}$ respectively. The solid lines are analytic predictions of $E_{\rm EPR}$ and of the spin lengths derived from a 4-mode model valid for a strictly adiabatic evolution (see text). The two (indistinguishable on the figure) dash-dotted lines in the bottom row represents the normalized density overlap (Eq.\eqref{eq:recouvrement}) of the wavefunctions $\phi_{a0}$ and $\phi_{a1}$ on the one hand and of $\phi_{b0}$ and $\phi_{b1}$ on the other hand.}
\label{fig:TINTIN}} 
\end{figure*}

\begin{figure}[htb]
$N_a=N_b=500$ \\
\includegraphics[width=0.47\textwidth]{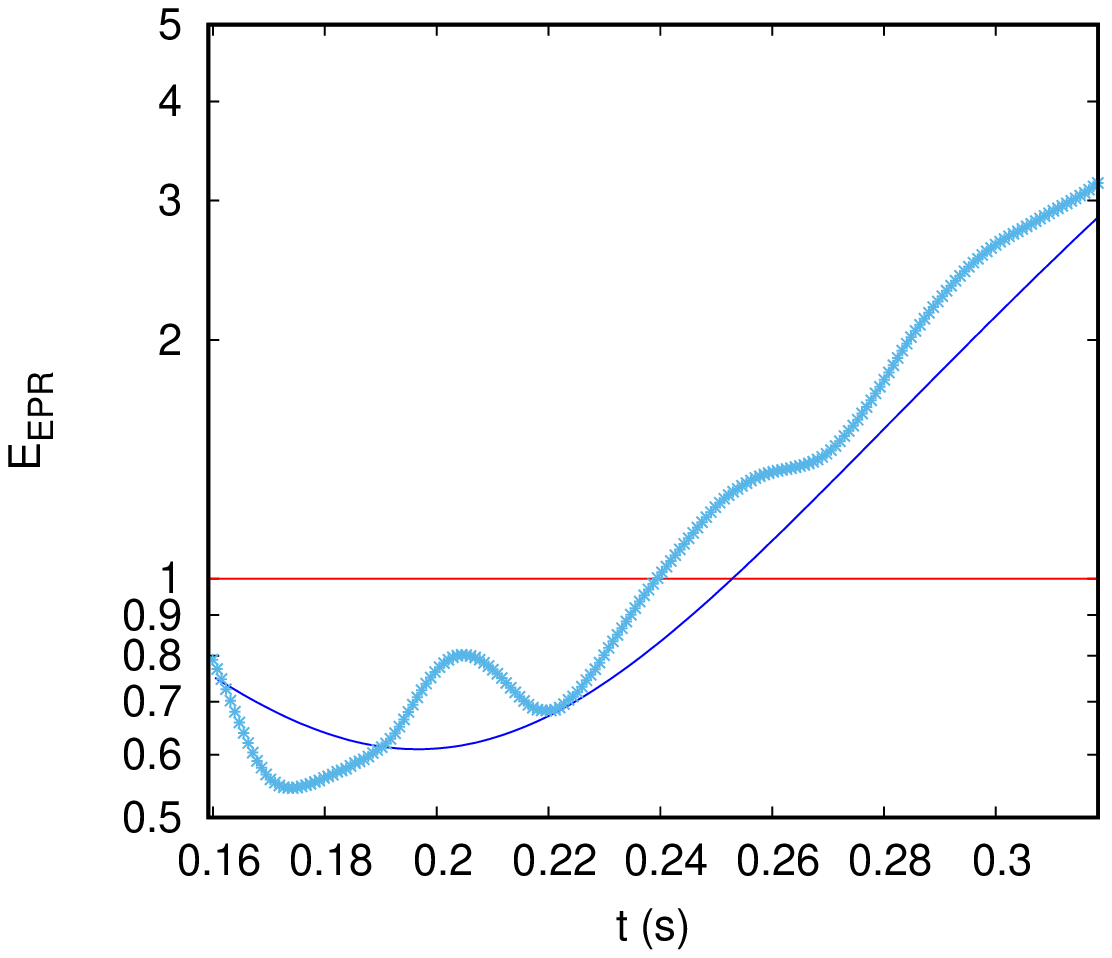} \\
\includegraphics[width=0.47\textwidth]{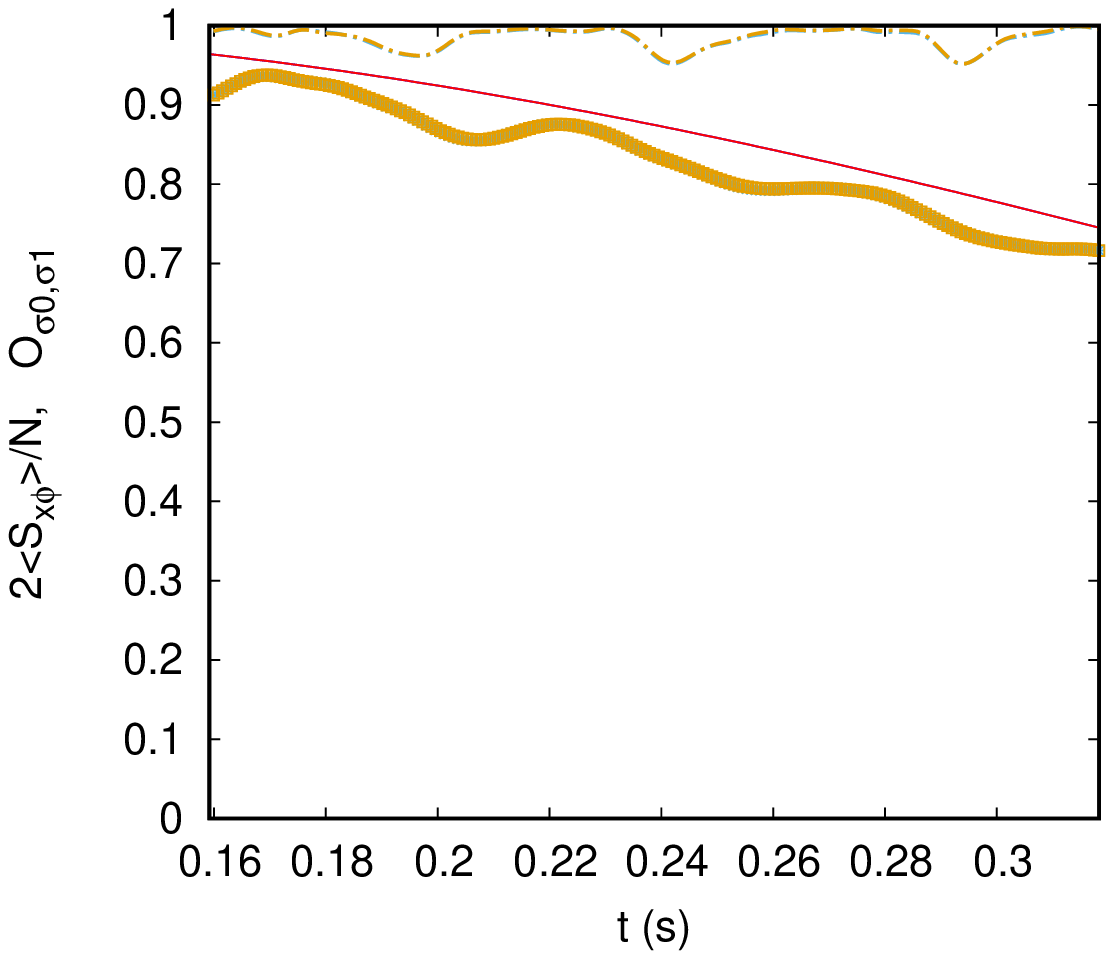}
\caption{(Color online) The EPR entanglement witness $E_{\rm EPR}$ (top) and the renormalized lengths of the two spins $2\meanv{\hat{S}_{x\phi}^a}/N_a$ and  $2\meanv{\hat{S}_{x\phi}^b}/N_b$ (bottom) are plotted as a function of the total evolution time $t=2t_{\rm R}+t_{\rm int}$ for $N_a=N_b=500$. The other simulation parameters are the same as in Fig.\ref{fig:TINTIN} except for a smaller initial separation between the traps $\delta z_{\rm max}=6 a_0$. Symbols: result of numerical simulations of spatial dynamics in the modulus-phase approximation (see Sec \ref{sec:modulephase}). \resoumis{In the top pannel, the blue crosses represent $E_{\rm EPR}$.} In the bottom pannel, the blue crosses and yellow squares are for $\meanv{\hat{S}_{x\phi}^a}$ and $\meanv{\hat{S}_{x\phi}^b}$ respectively. The solid lines are from an analytic 4-mode model valid for a strictly adiabatic evolution (see text). \resoumis{The two (indistinguishable on the figure) dash-dotted lines in the bottom row represents the normalized density overlap (Eq.\eqref{eq:recouvrement}) of the wavefunctions $\phi_{a0}$ and $\phi_{a1}$ on the one hand and of $\phi_{b0}$ and $\phi_{b1}$ on the other hand.}
\label{fig:SIX}} 
\end{figure}


\subsection{Experimental and numerical sequence}
\label{sec:sequence}

The experiment we simulated is inspired by the situation available on an atom chip \cite{Treutlein2010}, and is depicted on Fig.\ref{fig:sequence}. It is realized by the following sequence:

$(i)$ We populate a double well potential \revisionbis{with $N_a$ atoms of Rubidium $87$ in the
hyperfine state $\ket{F=1 , m_{\rm F}=-1}$}
in the many-body ground state of the \revisionbis{$a$} well and similarly in the \revisionbis{$b$} well with $N_b$ atoms. Different simulations were performed varying the initial trap distance and the atom number in the range $N_a=N_b=100$ to  $N_a=N_b=5000$. Each well is composed of an isotropic harmonic trap whose trapping frequency is fixed to
\be
\omega=2\pi\times 20\, \mbox{Hz}
\ee
The system is cylindrically symmetric around the axis of the two trap centers. We choose this axis as the $z$ axis of a cylindrical frame and write the spatial dependence of the wave functions as $\phi(\rr)=\phi(r,z)$.

$(ii)$ A $\pi/2$-pulse drives the atoms in both wells into a superposition of the two long-lived hyperfine states \revisionbis{$\ket{0}=\ket{F=1,m_F=-1}$ and $\ket{1}=\ket{F=2,m_F=1}$.
The scattering lengths characterizing the interactions between these hyperfine states are known experimentally within an accuracy of about $0.5\%$ \cite{Hall2007,Sidorov2013}, and depend on the external magnetic field \cite{Verhaar2002}. The results presented in this article are for the values
\resoumis{
\be
a_{00}=100.4\, R_{\rm B} \quad a_{11}= 95.0\, R_{\rm B} \quad a_{01}=98.0\, R_{\rm B}
\label{eq:longdiff}
\ee
in units of the Bohr radius $R_{\rm B} \simeq 52.918\times10^{-12}\, \mbox{m}$ \footnote{\resoumis{We checked that a relative change of the order of 0.5\% of the scattering lengths does not significantly affect the results in our scheme by performing test simulations for the sets of values $a_{00}=100.0\, R_{\rm B} \quad a_{11}= 95.0\, R_{\rm B} \quad a_{01}=98.0\, R_{\rm B}$ and \resoumis{$a_{00}=100.44\, R_{\rm B} \quad a_{11}= 95.47\, R_{\rm B} \quad a_{01}=98.28\, R_{\rm B}$} for $N_a=N_b=5000$.}}
}}.

$(iii)$ The trapping potential of atoms in state $\ket{0}$ is displaced by a distance exactly equal to the initial separation between the two wells using a ramp of fixed duration $\omega t_{\rm R} = 10$. In order to minimize the creation of spatial excitations, the ramp should be slow in the beginning and in the end, when the atomic clouds are being mixed or demixed, and faster in between. This is achieved by varying the displacement $\delta z$ of the trapping potential of state $1$ using an hyperbolic tangent function of time
\be
\frac{\delta z(t)}{\delta z_{\rm max}} = \frac{\mbox{th}(4t/t_{\rm R}-2)-\mbox{th}(-2)}{\mbox{th}(2)-\mbox{th}(-2)}
\label{eq:rampe}
\ee
where $\delta z_{\rm max}$ is the initial separation between the two wells and $t_{\rm R}$ is the duration of the ramp.

$(iv)$ After the end of the ramp, component 0 of the \revisionbis{$a$} well interacts with component 1 of the \revisionbis{$b$} well during an adjustable interaction time $t_{\rm int}$.

$(v)$  Components 0 are ramped back to their initial positions using the ramp \eqref{eq:rampe} backwards.

$(vi)$ Averages and correlations between spin components are computed from \resoumis{Eq.\eqref{eq:resultat_final}.}

$(vii)$ The minimal value of $E_{\rm EPR}$ is found by numerically optimizing over the angles $\alpha$ and $\beta$ (see Eqs.\eqref{eq:alpha} and \eqref{eq:beta}).

The main free parameters of this experiment are the initial spatial separation $\delta z_{\rm max}$ between the two wells and the ramping time $t_{\rm R}$. The separation $\delta z_{\rm max}$ should be large enough to prevent any overlap between the two wells, initially and after ramping back, thus ensuring the spatial separation required for the EPR experiment. The ramping time $t_{\rm R}$ should be long to avoid too violent excitations of the spatial dynamics of the clouds, which would degrade the creation of EPR correlations. There is also \revisionbis{an upper} bound on $t_{\rm R}$: during the ramp, correlations build up between intrawell spin components (\textit{e.g.} between $\hat{S}_{y\phi}^a$ and  $\hat{S}_z^a$) like in a spin-squeezing experiment \cite{TreutleinSinatra2009} ; these correlations are unfavorable for our purpose because they are a source of uncertainty for non local observables such as $\hat{S}_{y\phi}^a \hat{S}_z^b$. 


\subsection{Discussion of the results}
\label{sec:discussion}

As a first example, in Fig.\ref{fig:TINTIN} we test the effect of spatial dynamics on EPR correlations by considering a relatively fast transport of the atomic clouds with
$\delta z_{\rm max}=10 a_0$ in unit of the oscillatory length $a_0=\sqrt{\hbar/m \omega}$ and the ramping time $\omega t_{\rm R}=10$, and two atom numbers $N_a=N_b=100$ and $N_a=N_b=500$.
The results of the numerical simulation for $E_{\rm EPR}$ and for the renormalized spin length $\meanv{\hat{S}_{x\phi}^\sigma}$ {including} spatial dynamics in the modulus-phase approximation are given by symbols \resoumis{respectively in the top and bottom row of Figs.\ref{fig:TINTIN} and \ref{fig:SIX}}. \resoumis{In addition, we show the normalized density overlap 
\be
\mathcal{O}_{\sigma 0,\sigma 1}=\frac{\int \dd^3 r |\phi_{\sigma 0}|^2 |\phi_{\sigma 1}|^2}{\sqrt{\int \dd^3 r |\phi_{\sigma 0}|^4 \int \dd^3 r |\phi_{\sigma 1}|^4}}
\label{eq:recouvrement}
\ee
between wavefunctions $\phi_{\sigma 0}$ and $\phi_{\sigma 1}$, as dash-dotted lines in the bottom row.} \revision{Each point is obtained by a numerical simulation corresponding to a different interaction time, always including the initial phase in which the potentials are displaced and the final one in which they are ramped back to the initial position before the measurement.} We compare these results to an analytical 4-mode model represented in solid-line ; this model assumes that the whole evolution (the $\pi/2$-pulse and the onward and backward ramps) is adiabatic so that the 4 wave functions remain in their instantaneous ground state at all time. With this assumption, the system can be mapped onto a 4-mode model whose non-linearity parameters $\chi_a$, $\chi_b$ and $\chi_{ab}$, coefficients respectively of $(\hat{S}_z^a)^2$, $(\hat{S}_z^b)^2$ and $-\hat{S}_z^a\hat{S}_z^b$ in the 4-mode Hamiltonian \cite{KPS2013}, are time-dependent and given by
\bea
\!\!\!\!\!\!\!\!\!\!\!\!\chi_\sigma(t) &=& \frac{1}{2} \bb{\frac{\partial}{\partial N_{\sigma0}}-\frac{\partial}{\partial N_{\sigma1}}} (\mu_{\sigma0}-\mu_{\sigma1}), \ \sigma=a,b \\
\!\!\!\!\!\!\!\!\!\!\!\!\revisionbis{\chi_{ab}(t)} &=& \revisionbis{\frac{1}{2} \bb{\frac{\partial \mu_{b1}}{\partial N_{a0}}+\frac{\partial\mu_{a0}}{\partial N_{b1}}}}
\eea
where $\mu_\alpha$ is the ground state chemical potential of mode $\alpha$. \revisionbis{In practice, for the values \eqref{eq:longdiff} of the scattering lengths, we have $\chi_a \simeq \chi_b\equiv\chi$.} In the formulas of Appendix B of Ref.\cite{KPS2013}, {valid for a purely stationary 4-mode model,} we perform the substitutions $\chi t \to \int_{0}^t \chi(t') \dd t'$ and $\chi_{ab} t \to \int_{0}^t \chi_{ab}(t') \dd t' $, where $t=2t_{\rm R}+t_{\rm int}$ is the total evolution time. The optimized EPR entanglement witness obtained through this approach is shown on \resoumis{the top row of} Fig.\ref{fig:TINTIN} in solid line.

The analytical $4$-mode model captures well the general trend of variation of $E_{\rm EPR}$ and of the renormalized spin length $\meanv{\hat{S}_{x\phi}^\sigma}$. 
$E_{\rm EPR}$ reaches a minimum for an optimal time \cite{KPS2013}, which is the result of a competition between the simultaneous creation of local and non-local correlations and the loss of coherence. Notice that this minimum is lower for 500 atoms ($E_{\rm EPR}\simeq 0.65$) than for 100 ($E_{\rm EPR}\simeq 0.85$) (see also \cite{KPS2013}), indicating that the EPR inequality is violated more strongly for larger atom number.

On top of this trend, a strong spatial dynamics is visible: at low atom number, the dynamics is dominated by regular oscillations of the centers-of-mass of the wave functions in the first excited mode of the harmonic trap. This is mostly an ideal gas behavior, caused by the finite ramping time; the weak interactions manifest themselves in the synchronization of the oscillations of $\phi_{a0}$ and $\phi_{b1}$, caused by the weak coupling between them (see Supplementary Material). This dynamics is responsible for the regular drops in the spin length $\meanv{\hat{S}_{x\phi}^\sigma}$ visible on Fig.\ref{fig:TINTIN}.$(a)$ with a frequency that matches the trap frequency. \resoumis{Remark that the spin length is close to its maximal value only when the density overlap is close to one. However, the converse is not true because perfectly overlapping wavefunctions may still be dephased. Ultimately, the dips of the spin length} causes the regular peaks in the EPR entanglement witness. 
At large atom number $N_a=N_b=500$ (Fig.\ref{fig:TINTIN}.$(b)$) the spatial dynamics is both damped and rendered more chaotic by the stronger interactions, and the generation of non-local correlations is faster.

We found that the results are improved by reducing the initial distance between the two traps. This increases the ratio between the interaction time and the total time and minimizes the excitation of spatial dynamics. We show an example for $\delta z_{\rm max}=6 a_0$ and $N_a=N_b=500$ in Fig.\ref{fig:SIX} where the minimum value of $E_{\rm EPR}\simeq 0.54$ is reached at $t\simeq0.17$ s. The corresponding plot for $\meanv{\hat{S}_{x\phi}^\sigma}$ is shown in the lower panel of Fig.\ref{fig:SIX}. A slight improvement ($E_{\rm EPR}\simeq 0.45$ \resoumis{at $t\simeq 0.16\,$s}) is obtained when increasing the number of atoms to $N_a=N_b=5000$ but the spatial dynamics is more strongly excited. 

\resoumis{To complete our investigation, we have evaluated the total loss rate in the interaction configuration (middle row in Fig.\ref{fig:sequence}) using stationary wave functions for $N_a = N_b = 500$, scattering lengths and trap frequencies as in Fig.\ref{fig:TINTIN}, and the loss rate constants \cite{SidorovDrummond2012} $\kappa_{11}=81\times10^{-21} \,\mbox{m}^{3}\ \mbox{s}^{-1}$, $\kappa_{01}=15\times10^{-21} \,\mbox{m}^{3}\ \mbox{s}^{-1}$ for two-body losses and $\kappa_{000}=5.4\times10^{-42} \,\mbox{m}^{6}\ \mbox{s}^{-1}$ for three-body losses. We find that about $7$ particles are lost on average in $0.2$ s  due to two-body losses, while three-body losses are negligible. If we add one-body losses corresponding to a lifetime of $1$ min, we obtain a total lost fraction $N_{\rm lost}/(N_a+N_b)$ of about  $10^{-2}$ at $0.2$ s. In analogy with what happens for spin-squeezing, where the losses are negligible as long as the lost fraction is much smaller than the squeezing parameter $\xi^2$ (see Eq.(21) in Ref.\cite{YunSinatra2008}), we expect that the effect of losses should not significantly affect the predicted value of $E_{\rm EPR}$ for the considered parameters. We come to a similar conclusion for larger atom numbers $N_a = N_b = 5000$ where we estimate a lost fraction of $2\times10^{-2}$ at the optimal time for EPR steering $t\simeq0.16\,$s.}

\resoumis{Our results show that an EPR steering experiment can be performed} using interactions in Bose-Einstein condensates in state-dependent potential even accounting for the spatial dynamics. This completes our previous study \cite{KPS2013} which was limited to stationary modes.

\section*{Conclusion}
We studied a system of Bose-Einstein Condensates in 4 distinguishable components entangled by collective transport in state-dependent potentials. Far beyond a simple 4-mode approach, we worked out a realistic description of the spatial structure of the system where both the spatial excitations and the entanglement between the distinguishable components are accounted for. 
From the methodological point of view, we generalized the two-component spatial dynamical model of Refs.\cite{CastinSinatra2000,TreutleinSinatra2009} to a many-component system, and we worked out a general formula to compute all the relevant correlations. 
{Our compact formulation makes our theory an easily adaptable tool to study the effect of spatial dynamics on entanglement, and provides the experimentalists with a numerical recipe to benchmark their results.}

Based on this description, we performed a set of numerical simulations to investigate the effects of spatial dynamics, and in particular of the excitations created by the collective transport of atomic clouds, on the non-local correlations which we expect in our system.
We observed that strong excitations of the spatial modes were created, whose behavior became more chaotic as the number of atoms increased. However, despite these excitations, the non-local correlations remain strong enough to perform an \resoumis{EPR steering experiment}. If the parameters, and in particular of the collective transport, are properly adjusted, our EPR entanglement witness can reach below the critical value of $1$, and down to $E_{\rm EPR}=0.54$ for $500$ atoms in each well and $0.45$ for $5000$. The lowest values of our entanglement witness were reached at relatively short times, of the order of \resoumis{$0.2$\, s.} {This is longer than the times ($\approx 15$\, ms) at which spin squeezing was observed with Bose Einstein condensates \cite{Treutlein2010} but remains within reach of cold atomic experiments \cite{Dalibard2002}.}


\section*{Acknowledgments}
\revisionbis{
H. Kurkjian is a FWO [PEGASUS]\textsuperscript{2} Marie Sk\l odowska-Curie fellow.
This project has received funding from the FWO and the European Union's Horizon 2020 research and innovation program under the Marie Sk\l odowska-Curie grant agreement number 665501. 
K. Paw\l owski acknowledges support by the (Polish) National Science Center
Grant No. 2014/13/D/ST2/01883.}


\providecommand*\hyphen{-}

\end{document}